\documentclass[aip,jmp,reprint]{revtex4-1}

\draft 

\usepackage{graphicx}
\usepackage{color}
\usepackage{float}
\begin{document}


\title{All-optical switching via four-wave mixing Bragg scattering in a silicon platform} 



\author{Yun Zhao}
\affiliation{University of Dayton, Department of Electro-Optics, Dayton Ohio 45469}
\author{David Lombardo}
\affiliation{University of Dayton, Department of Electro-Optics, Dayton Ohio 45469}
\author{Jay Mathews}
\affiliation{University of Dayton, Department of Electro-Optics, Dayton Ohio 45469}
\affiliation{University of Dayton, Department of Physics, Dayton Ohio 45469}
\author{Imad Agha}
\affiliation{University of Dayton, Department of Electro-Optics, Dayton Ohio 45469}
\affiliation{University of Dayton, Department of Physics, Dayton Ohio 45469}
\email[iagha1@udayton.edu]{email}


\date{\today}

\begin{abstract}
We employ the process of non-degenerate four-wave mixing Bragg scattering (FWM-BS) to demonstrate all-optical control in a silicon platform. In our configuration, a strong, non-information-carrying pump is mixed with a weak control pump and an input signal in a silicon-on-insulator waveguide. Through the optical nonlinearity of this highly-confining waveguide, the weak  pump controls the wavelength conversion process from the signal to an idler, leading to a controlled depletion of the signal. The strong pump, on the other hand, plays the role of a constant bias. In this work, we show experimentally that it is possible to implement this low-power switching technique as a first step towards universal optical logic gates, and test the performance with random binary data. Even at very low powers, where the signal  and control pump levels are almost equal, the eye-diagrams remain open, indicating a successful operation of the logic gates.     
\end{abstract}


\maketitle 

All-optical control of light on an integrated chip is essential for the implementation of photonic circuits for signal processing and computing applications. In recent years, great effort has been placed on implementing on-chip all-optical switching using  various nonlinear phenomena. Essentially, the idea is to control one beam of light with another, i.e. to switch it in a controlled manner akin to that employed in modern electronics. Physical phenomena employed in previous work include free carrier dispersion \cite{Almeida2004}, cross absorption modulation \cite{Liang2005}, free carrier effect on the $\chi^{(3)}$ coefficient\cite{Hache2000}, active plasmonics \cite{Macdonald2009} and modulation of the magnetic resonance \cite{Shcherbakov2015}.  However, in most of those  experiments, the control beam is required to have Watt-level peak power, making it incompatible with many telecommunications components,  and placing ``synchronization'' requirements on an amplified laser. Moreover, the reliance on absorption to achieve signal control  also results in a considerable amount of heat generation, leading to an increase in energy consumption.   Additionally, it may lead to performance degradation over time, making such devices inferior to electronic switches in the very aspect in which they are supposed to surpass their electronic counterparts \cite{Biberman2012}. \\

\begin{figure}
\includegraphics[width=\linewidth]{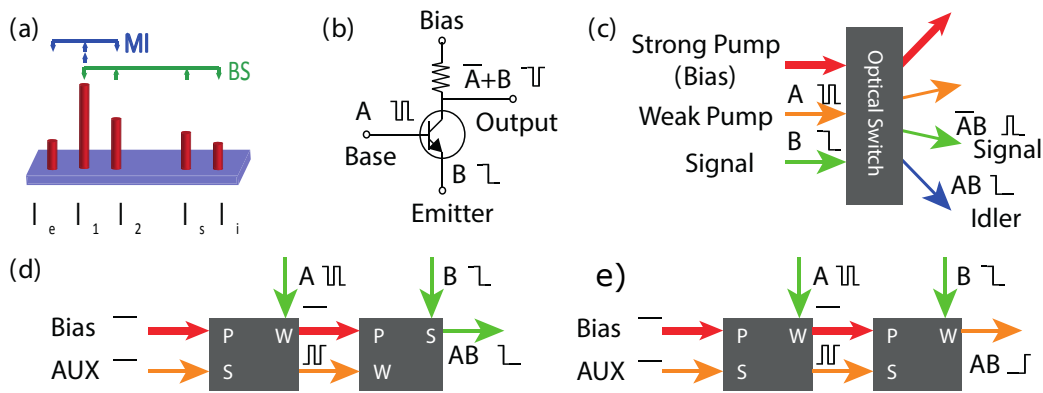}%
\caption{\label{Figure1} (a) Demonstration of two four wave mixing processes. MI, modulation instability; BS, Bragg scattering. (b) One common configuration of transistors in digital circuits. (c) The logic operation of an optical switch based one FWM-BS. (d) Two-stage configuration to create an AND gate. (e) Two stages to create a NAND gate. In the diagrams, the different wavelengths are color-coded. For example, in our current implementation, (c), orange=A=weak pump, green=B=signal, red=bias=strong pump, and the output  $\bar{A}B$ is read at the signal wavelength. }%
\end{figure}

In this work, we take the approach of using optically-controlled wavelength conversion in silicon to realize a low-switching power, low-heat generation and low-noise all optical switch. The process of interest is four wave mixing Bragg scattering (FWM-BS), a subcategory of FWM, where two nondegenerate pumps are on the same side (spectral domain) of the signal and the idler (Fig \ref{Figure1}a). The process converts photons from the signal wavelength to the idler wavelength without excess noise \cite{McKinstrie2005, Agha13}. Earlier work \cite{Agha2012,Bell2016} in FWM-BS tended to use two equally strong pumps to both  relax phase matching requirements and ensure high conversion efficiency. In a recent experiment, we showed that it is  possible to set one pump to a relatively weak power and compensate for the efficiency drop by increasing the other pump \cite{Zhao2016}. From an information processing perspective, this configuration is similar to a commonly used configuration of electronic transistors (Fig. \ref{Figure1}b,c). The strong optical pump is analogous to the electrical bias voltage, which enables the function of the device but does not interfere with the information flow. In such an electronic device, the base binary signal controls the flow of the emitter binary data to the output port, implementing the logic operation  $\bar{A}B$. In our configuration, the FWM-BS weak pump works as the base input and the FWM-BS signal carries the emitter binaries. The output (either FWM-BS signal or FWM-BS idler) binaries are the signal binaries controlled/modulated by the weak pump. We stress the importance of this approach, as the weak pump is in the range of power that is safe for standard telecommunications components (non-amplified lasers, modulators, wavelength-division multiplexers, filters, etc...), and, as such, easily carries all the synchronization and timing information. On the other hand, the strong pump - intrinsically incompatible with such devices- can bypass them altogether. As an example, we implement the $\bar{A}B$ gate, which is easily achievable in a one-pass configuration. We note here that while the one-pass configuration creates the $\bar{A}B$ gate, by harnessing the FAN-OUT capability of our device, it is possible to create a universal optical gate. Fig. \ref{Figure1}d shows how two gates back to back can be used to make an AND gate, while (Fig. \ref{Figure1}e) shows how a NAND gate can be implemented. \\


As shown in Fig \ref{Figure1}, FWM-BS involves 4 waves satisfying the energy conservation relation,  
\begin{equation}
\frac{1}{\lambda_1} + \frac{1}{\lambda_i} = \frac{1}{\lambda_2} + \frac{1}{\lambda_s}
\label{energy}
\end{equation}
or,
\begin{equation}
\frac{1}{\lambda_1} + \frac{1}{\lambda_s} = \frac{1}{\lambda_2} + \frac{1}{\lambda_i},
\label{energy2}
\end{equation}

where $\lambda_1, \lambda_2, \lambda_s, \lambda_i$ are wavelengths of two pumps, the signal and the idler respectively. It is  desirable in this configuration to suppress the degenerate four wave mixing (FWM-MI) centered at $\lambda_1$ since it introduces amplification-related excess noise. This can be done by controlling the chromatic dispersion in a guided wave structure. If we assume pump 1 is the only undepleted pump, the phase matching condition for narrow-band FWM-BS can be worked out, using a similar method\cite{Chen1989} to  earlier theoretical work, to be:
\begin{equation}
\gamma P_1 = \Delta\beta \approx -\frac{2\pi c}{\lambda_0^2}\left.D\right|_{\lambda_0}(\lambda_1-\lambda_2)(\lambda_1-\lambda_s)
\label{PhaseMatching1}
\end{equation}
or
\begin{equation}
\gamma P_1 = \Delta\beta \approx -\frac{2\pi c}{\lambda_0^2}\left.D\right|_{\lambda_0}(\lambda_1-\lambda_2)(\lambda_1-\lambda_i)
\label{PhaseMatching2}
\end{equation}
where $\gamma$ is the effective nonlinear coefficient for the waveguide cross-section, $P_1$ the power of the strong pump, $D$ the group velocity dispersion (GVD) coefficient and $\lambda_0$ the center wavelength. The phase matching condition for FWM-MI from pump 1 is \cite{Agrawal2007book}
\begin{equation}
\gamma P_1 = -\Delta\beta_{\text{MI}} \approx \frac{2\pi c}{\lambda_1^2}\left.D\right|_{\lambda_1}(\lambda_1-\lambda_s)^2.
\end{equation}
We see that the phase matching of FWM-MI can only be satisfied when $D$ is positive while FWM-BS can be phase matched even when $D$ is negative, assuming that the strong pump (pump 1) corresponds to the highest or lowest wavelength. This is significant since it implies that the nearly noise-free FWM-BS can be employed in a regime where the noisy FWM-MI process is inhibited. 

Silicon-on-insulator (SOI) is chosen as the platform of the experiment due to its compatibility with standard processing techniques and its strong optical nonlinearity \cite{Foster07}. Simple rectangular buried silicon waveguides are fabricated and the dimensions are chosen based on numerical simulations to match the desired dispersion that optimizes FWM-BS and inhibits FWM-MI. The waveguides are fabricated through standard techniques, including electron beam lithography, inductively coupled plasma etching (Cl$_2$ based) and plasma enhanced chemical vapor deposition (SiO$_2$). Resist reflow is applied after lithography to reduce side wall roughness. The etched waveguide size is 500 nm $\times$ 800 nm $\times$ 1.6cm with  3 $\mu$m wide adiabatic tapers on both ends to improve fiber to waveguide mode matching. The final devices have $\mathtt{\sim}$ 7 dB/facet coupling loss and  $<$ 3.5 dB/cm propagation loss.\\

\begin{figure}
\includegraphics[width=\linewidth]{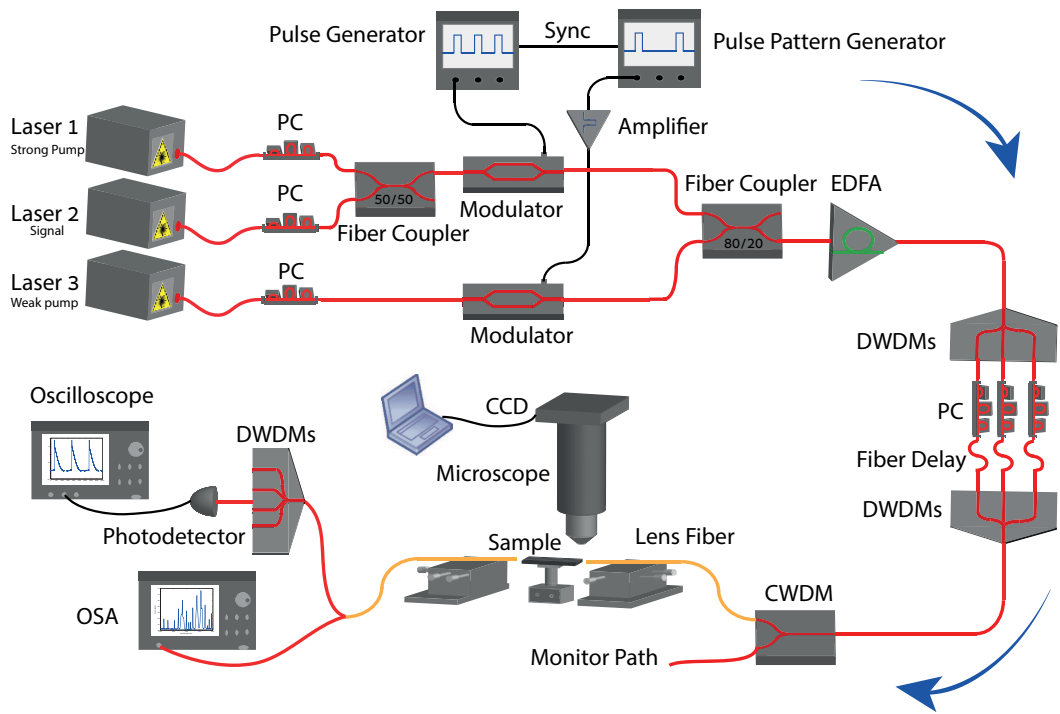}%
\caption{\label{FWMSetup} Experiment setup. PC, polarization controller.}%
\end{figure}

The experimental setup is shown in Fig \ref{FWMSetup} and implements the logic operation $\bar{A}B$ (which can be implemented in a single gate operation). Three telecom C-band lasers are used as the seeds for the strong pump, weak pump and signal respectively. The strong pump and the signal are modulated together into 1/128 duty cycle periodic pulses. The weak pump is encoded with pseudo random bits by a pulse pattern generator (PPG) that is synchronized to the periodic pulses. The random bits are originally generated by a computer, reshaped into the same duty cycle as the periodic pulses and then passed to the PPG through the general purpose interface bus (GPIB) communication port. The reason for the use of pulses is to boost the peak power of the strong pump during the amplification stage, where a saturated erbium doped fiber amplifier (EDFA) is used. The other lasers also go through the EDFA because some power boost is needed to compensate for the fiber to silicon coupling loss. This may raise the concern of an optical gain redistribution between high and low logic periods. In other words, the signal may experience more gain at the time when the weak pump is absent because of a weaker gain competition, which also resembles a switching behavior. However, this is insignificant because the pulse period is much smaller than the life time of erbium $^4I_{13/2}$ level; the random absence of the weak pump is not distinguishable by the erbium atoms \cite{Nilsson1993}. This is also verified by looking at the input signal which showed no power fluctuation beyond the intrinsic noise. Moreover, the modulation of the strong pump and the amplification of weak pump can possibly be avoided if a higher efficiency coupling scheme, such as inverse tapers \cite{Wood2012}, is implemented.\\
\indent As a significant amount of amplified spontaneous emission (ASE) and stimulated Raman scattering (SRS) noise are introduced into the optical signal after the amplification, multiple dense wavelength division multiplexers (DWDMs) with more than 30 dB isolation are used to clean up the spectrum. Each  wavelength channel of interest is assigned a separate  polarization controller to promote mode matching during the fiber-to-silicon coupling. The optical path lengths are carefully controlled so that all the pulses stay overlapped after recombining at the chip's input. The 980 nm port of a 1550/980 coarse wavelength division multiplexer is used as the monitor port for the input signals ($\mathtt{\sim} 1\%$ of the input power is diverted to the 980 nm port). A conical lens fiber converts the 10 $\mu$m fiber mode into a 2 $\mu$m free space spot, which is then launched into the silicon waveguide. Another lens fiber is placed at the output side of the waveguide to collect the output signal. DWDMs are again used to pick up the channel to be measured. A photodetector with variable gain and bandwidth is used to detect and show the signal on a digital sampling oscilloscope with 5 Gs/s sampling rate and 20 million samples per recording. The final results are then transferred onto a computer and analyzed.\\

\begin{figure}
\includegraphics[width=\linewidth]{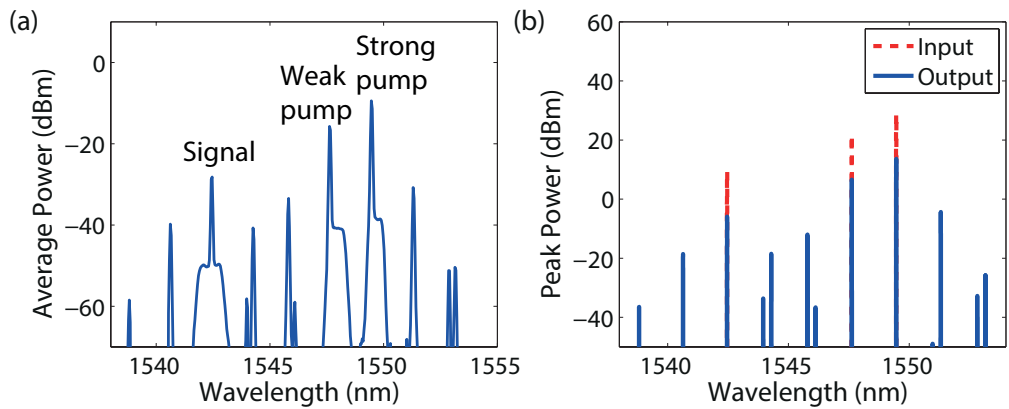}%
\caption{\label{Spectrum} (a) Experiment spectrum. (b) Simulated spectrum by split-step Fourier method.}
\end{figure}

First, the pulse pattern generator is set to generate periodic pulses and the optical spectrum is measured. A maximum of 2 dB wavelength dependent coupling efficiency difference is measured with a small signal input. Taking that into consideration, the strong pump power at the input of the waveguide is  set to 650 mW (peak power), the weak pump to 150 mW and the signal to 7.5 mW. 15 \% of the power is converted to the first order idlers. The split-step Fourier method with both real (FWM) and imaginary (two photon absorption) part of $\chi^{(3)}$ coefficients is also used to estimate the efficiency of the device numerically. The experimental efficiency (Fig. \ref{Spectrum}a) is comparable to the  simulated efficiency (Fig. \ref{Spectrum}b) when using parameters from recently published data on the nonlinearity and absorption in crystalline silicon\cite{Bristow2007,Weber2002}.\\

\begin{figure}[H]
\includegraphics[width=\linewidth]{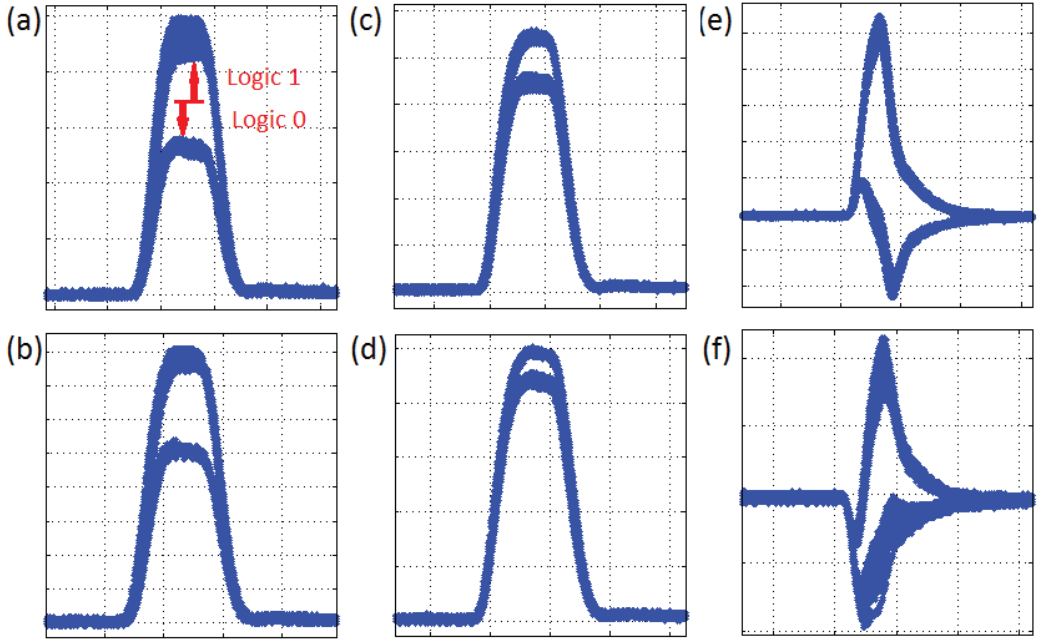}%
\caption{\label{Eye1} Eye-like diagrams of the FWM-BS signal. The strong pump is at 630 mW (peak power) for (a) and 680 mW for the rest. The signal is at 41 mW. (a) Weak pump at 360 mW. (b) Weak pump at 208 mW. (c) Weak pump at 100 mW. (d) Weak pump at 47 mW. (e) Weak pump at 100 mW with balanced detection. (f) weak pump at 47 mW with balanced detection.}%
\end{figure}

To verify  the switching capability of the device, we first consider the FWM-BS signal at the output and look for signal depletion as a function of the control pump. The quality of the switching operation is evaluated by plotting the eye-like diagrams (ED) of the output signal (due to the imperfect extinction of the signal, the baseline does not go to zero, which leads to the elevated floor in Fig. 4). Given the use of on/off pulses, the final diagram resembles that of a return-to-zero (RZ) signal. Fig \ref{Eye1} shows the center part of the generated eye diagrams. The strong pump is kept at around 680 mW peak power and the signal is at 41 mW. The distortion of the pulses, which happens after the EDFA amplification stage, significantly increases the original signal noise. When the weak pump is at 360 mW, approximately 50 \% signal extinction ratio is measured (Fig \ref{Eye1}a). The carried information can be easily retrieved without error. It is worth pointing out that this extinction ratio is a combined effect of FWM-BS and nondegenerate two photon absorption since the ``weak'' pump is also relatively strong. The noisy upper lid is due to the high power of the ``weak'' pump which causes an observable gain competition inside EDFA. When the power of the weak pump is lowered, the signal extinction ratio decreases as expected, with the FWM-BS contribution decreasing linearly and the TPA contribution decreasing nearly quadratically (Fig \ref{Eye1}b,c,d). However, since the FWM-BS process itself is noise free, the eye diagram remains open for a weak pump as low as 47 mW - almost the same as the signal power of 41 mW - indicating an operation with equal signal and control powers. In fact the upper lid is significantly cleaner at a lower control power due to a lower noise on the weak (control) pump.  The distinction of logic highs and lows can be further improved by keeping a reference beam and performing a standard balanced detection at the output (Fig \ref{Eye1}e,f). The non-flat lower lid in Fig \ref{Eye1}e,f is caused by a 2 ns optical path mismatch between the signal and the reference beam. This can be easily addressed by using a higher-precision mechanism for path matching, such as a tunable delay line.\\

\begin{figure}[H]
\includegraphics[width=\linewidth]{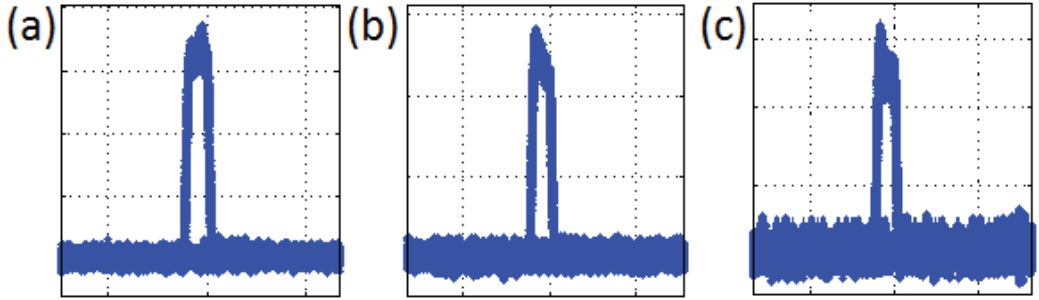}%
\caption{\label{Eye2} Eye diagrams of Idler. The strong pump is at 680 mW (peak power). (a) Weak pump at 100 mW. (b) Weak pump at 47 mW. (c) Weak pump at 25 mW.}%
\end{figure}

A significantly higher extinction ratio can be achieved by selecting the FWM-BS idler as the output signal \cite{AFoster16}, under which the logic low is automatically zero. As shown in Fig \ref{Eye2}, even when the weak pump is lower than the signal (25 mW vs. 41 mW), the ONs and OFFs of the idler can be easily distinguished (Fig \ref{Eye2}c). However, this scheme requires a shift in the output wavelength, and hence is not as conducive to multiple sequential gate operation (although this can be in principle rectified by flipping the role of signal and idler at each operation, or by employing a stage of wavelength conversion in between). \\

In conclusion, we have demonstrated a low control power, low-noise  optically controlled switch, based on four-wave mixing Bragg scattering, and implemented the logic gate $\bar{A}B$. Eye-diagram measurements show that the switch has low error rates,  owing to the nature of the nearly noise-free wavelength translation process. Future work will focus on higher data rates, universal gate demonstration, and multi-stage combinatorial logic operation.


%
%

%


\bibliography{YZ_bib_2016}

\end{document}